# USING UWB FOR HUMAN TRAJECTORY EXTRACTION

**Gonçalo Vasconcelos[2], Marcelo Petry[†,3], João Emílio Almeida[†,1], Rosaldo J. F. Rossetti[†,1], António Leça Coelho[4]**

[†] Artificial Intelligence and Computer Science Laboratory (LIACC)
[1] Department of Informatics Engineering (DEI)
[2] Department of Electrical and Computer Engineering (DEEC)
Faculty of Engineering, University of Porto (FEUP)
[3] INESC TEC – INESC Technology and Science

[4] Laboratório Nacional de Engenharia Civil (LNEC)

{ee03120, marcelo.petry, joao.emilio.almeida, rossetti}@fe.up.pt, alcoelho@lnec.pt

**ABSTRACT**
In this paper we report on a methodology to model pedestrian behaviours whilst aggregate variables are concerned, with potential applications to different situations, such as evacuating a building in emergency events. The approach consists of using UWB (ultra-wide band) based data collection to characterise behaviour in specific scenarios. From a number of experiments carried out, we detail the single-file scenario to demonstrate the ability of this approach to represent macroscopic characteristics of the pedestrian flow. Results are discussed and we can conclude that UWB-based data collection shows great potential and suitability for human trajectory extraction, when compared to other traditional approaches.

Keywords: UWB, pedestrian behaviour modelling, emergency planning, evacuation.

## 1. INTRODUCTION

The safety of people in emergency situations where panic can lead to unexpected or anti-social behaviour has been a subject of interest and study of researchers, engineers and authorities (Almeida 2011). The study of pedestrian motion and behaviour is thus crucial for assuring the safety of people in situations such as confined spaces with large concentrations of people, which can lead to a large number of wounded and dead people in an evacuation scenario. However, studying crowd behaviour in emergency situations is difficult since it often requires exposing real people to the actual, possibly dangerous, environments (Almeida 2012).

Although models and simulation tools exist to help building designers and emergency planners, results are not generally satisfactory. Scientists and practitioners most of the time must rely on simplified models for which validation is problematic due to the lack of appropriate data, as well as reference models and benchmarks with which to compare. These models must, therefore, be calibrated against reliable data to ensure their validity.

To improve the knowledge related to pedestrian dynamics, it is necessary to develop experiments with volunteers where their movements are recorded and data is subsequently mined in order to extract meaningful information. Towards this goal, this paper reports on a series of experiments performed in different scenarios, using a radio frequency based indoor localization to collect the trajectories of every participant. The data resulting from the experiments will be used in the context of a larger project: "mSPEED" Simulator, an integrated framework constituting a unique tool for agent-based "Modelling and Simulation of Pedestrian Emergent Evacuation Dynamics" under development at LIACC. This tool will be used to validate new and existing building layouts, help planners develop and improve emergency plans and safety systems, train occupants using virtual drills, and help fire-fighters and rescuers develop plans, rescue strategies, and learn how to deal effectively with crowds during emergencies and critical situations.

The remainder of this paper is organised as follows. Section 2 presents some pedestrian data collection technologies, whereas Section 3 is used to detail the UWB-based approach for human trajectory extraction. Experiments are described in Section 4, whose results are discussed in Section 5. We finally draw conclusions and point out directions for further research in Section 6.

## 2. PEDESTRIAN DATA COLLECTION

Crowds and pedestrians have been empirically studied for the past decades (Cordeiro 2011; Predtechenskii and Milinskiiand 1978). The evaluation methods applied were based on direct observation, photographs, and time-lapse films. Video has been an important tool for observing pedestrians (Helbing 2001). It is considerably better than direct observation, since the same scene can be studied over and over again, by human observation or using computer software with analysis algorithms.

Some research laboratories have been using video recordings of pedestrians to study and analyse their movement and extract patterns that can provide insight into theories and mathematical models. One example is the Pedestrian Accessibility Movement Environment Laboratory – PAMELA built to better understand how people interact with the environment and thus how can designers mitigate some of the problems people encounter as they move around (http://www.ucl.ac.uk/arg/pamela). Its aim is to study and develop tools for helping people with accessibility problems or some kind of motor disability.

Besides video, other technologies are also used to capture pedestrian behaviour, such as Bluetooth, RFID (Sharma and Gilfford 2005), among others. One possible way of addressing this challenge is by using a combination of various sensors, called "Sensor Fusion", combining data gathered from various sources, to help understand pedestrian movement and behaviour and thus enabling the creation of better and more complete databases.

In this paper, we propose the use of an ultra-wideband (UWB) radio frequency based indoor localization system for pedestrian dynamics trajectory tracking.

## 3. USING UWB FOR HUMAN TRAJECTORY EXTRACTION

### 3.1. UWB
According to Nekoogar (2005), the usage of UWB based technology offers several advantages over narrowband communication systems, some of which Corrales (2008) claim are beneficial for indoor human tracking:

- Due to the short duration of UWB pulses, receivers are able to differentiate the original signals from the reflected and refracted ones, which make UWB based systems less sensitive to multipath fading.
- The low power UWB signals reside below the noise floor of typical narrowband receivers and enable UWB signals to share the frequency spectrum with other radio services with minimal or no interference.
- There are no line-of-sight restrictions due to the long wavelength, low frequencies included in the broad range of the UWB frequency spectrum ability to penetrate a variety of materials.
- As UWB transmission is carrierless, no modulation is required, and the low-powered pulses eliminate the need for a power amplifier, resulting in simple transceiver architecture and reduced infrastructure.

Although the experiments presented in this paper are very simple and performed under laboratory conditions, the usage of a radio frequency based system allows for a number of advantages over traditional data collection methods like automatic extraction of pedestrian trajectories from video recordings: as previously mentioned, there are no line-of-sight restrictions and is also suited for low ceiling buildings. It also encompasses a wider breath of simulation scenarios, such as limited visibility situations (e.g. dark or smoke filled rooms), which are very common in emergency situations due to fire.

By assigning identifiable tags to individual participants, we are also able to easily associate some characteristics of the individual pedestrian (e.g. gender, height and age) with its trajectory. This might allow better understanding of the dynamics of heterogeneous crowds and study the effect of outliers like elderly people or people with mobility impairments.

### 3.2. Ubisense Description
The Ubisense (http://www.ubisense.net/) real-time location system is an in-building ultra-wideband radio based tracking system which can obtain accurate information of the positions of people and objects. This system uses small devices (tags) that send UWB pulses to a network of hardware receivers fixated in the localization area, which use a combination of TDOA (Time-Difference of Arrival) and AOA (Angle of Arrival) techniques to estimate the position of each tag (Steggles and Gschwind 2005). These tags can be attached to objects or carried by personnel. Sensors can also be connected to a computer, and Ubisense also provides a middleware platform which can manage and filter real-time location information and simplify the creation of location aware applications that monitor several localization areas simultaneously.

The system provided by Ubisense has seen wide adoption by the industry, especially in manufacturing plants, providing location services for tracking assets in order to improve and better control processes (Cadman 2003). Lately, it has also been used to track personnel during military and fire fighter training and operations (O'Conner 2005), and as a behaviour analysis tool based on coordinates of body tags (Luštrek 2009).

## 4. EXPERIMENTS

### 4.1. Aim
Our aim for the experiments described in this document is twofold. First, we aim to provide valid data for pedestrian dynamics model elicitation, as well as model validation in different facilities. Another goal is to evaluate the usage of a UWB based real-time location system for pedestrian movement data collection and trajectory extraction.

### 4.2. Setup
Experiments were performed in our lab, at FEUP (Fig.1), and were conducted with up to 30 participants, mostly students, whose average age was 21.4 ± 4.5 years and of mixed gender. For all experiments, the test persons were instructed to move with purpose, but not haste. These experiments follow a set of similar experiments performed in Germany at the Jülich

Supercomputing Centre (JSC) (Seyfried 2005, Seyfried 2007, Seyfried 2008, Boltes 2011), in which pedestrian trajectories were accurately extracted from video recordings. In contrast, in our experiments automatic data collection is performed by assigning individual tags to participants, whose position was then tracked. Due to the limited number of tags available, our experiments were performed on a smaller scale.

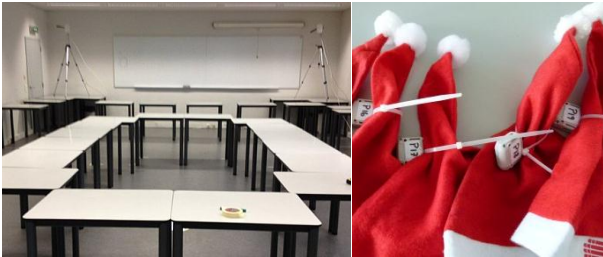

Figure 1: Experiment setup a) Room and sensors b) Christmas hats with tags

After being secured at a height of 2.3 m from the ground, four sensors were placed in the four vertices of the bounding rectangle containing the area where experiments took place (Fig.1a). Within this area, different scenarios were built using tables and a vinyl foldable wall.

Tag placement plays an important role in ensuring that good readings will be achievable. Tags become difficult to read when they are in close proximity to materials that absorb a large amount of radio frequency energy, such as water, that makes up most of the human body. For the purposes of these experiments, tags were attached on top of Christmas hats secured with straps (Fig.1b). Although the system works well without direct line-of-sight, placing tags this way ensures unobstructed path between tags and sensors, which improves measurements and is advised by the Ubisense system setup instructions.

The experiments were divided in three general set-ups: single-file, narrow passage and corner and T-junction. Each set-up aims to provide data suitable to study different phenomena: fundamental relation in a simple scenario, unidirectional pedestrian flow through bottlenecks and more complex configurations like corners and merging of flows respectively. In this paper we are concerned to analyse the single-file scenario only, which is described in detail in the following sections.

### 4.3. The Single-File Scenario

The reduced degrees of freedom associated with movement along a line helps minimize the effects that might influence the relation between the density and velocity of pedestrian movement. Single-file movement is therefore the simplest system for investigation of this dependency, which is the goal of the first experiment. To avoid boundary effects and limit the amount of test people needed, this first scenario was composed of a looping track. Corridor width was defined such that it does not impede the free movement of arms but enforces single-file movement by preventing passing and the formation of multiple lanes (Fig.2).

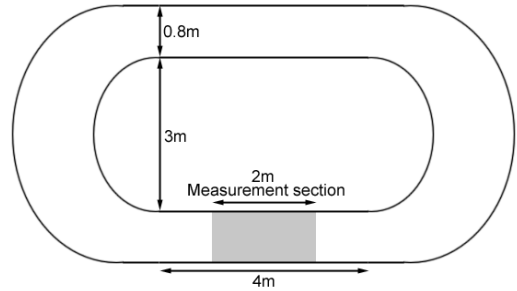

Figure 2: Experimental setup for the single-file scenario

Test subjects were instructed not to overtake and were distributed uniformly in the corridor. Each experiment run had each person complete two full loops before leaving the track. To regulate pedestrian density, three runs with 10, 15 and 20 randomly chosen pedestrians were performed. Ten runs with a single pedestrian were also performed with the purpose of free velocity determination.

Although trajectory data was collected for the whole track, only a section of the straight part with a length of 2 m was analysed, to avoid the influence of curves (shaded area in Fig.2).

### 5. RESULTS AND DISCUSSION

The UWB system used for data collection asynchronously reads tags' location: a stream of location events is generated over time; each event only contains information about the position of a single tag. Consequently the location of the crowd is updated one pedestrian at a time. Moreover, the system does not ensure a constant frequency of readings for each tag.

The mean frequency of location updates for a single tag in the collected data was $4.74 \pm 1.74$ Hz. This compares unfavourably with video collection techniques, where frequencies of 25 Hz are common and each frame contains data about all pedestrians.

One of the tags used during the experiments behaved erratically, and therefore was not considered for the purpose of representing trajectories. From the remaining stream of location events, trajectories were constructed for each pedestrian, whose representation for some experiments is presented in Fig.5.

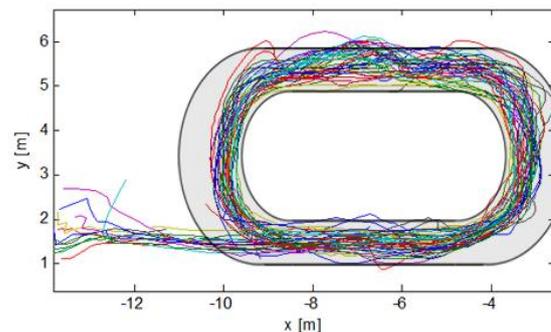

Figure 5: Extracted trajectories for the single-file scenario.

The collected trajectories appear to be jerky and imprecise. Considerable noise seems to affect some measurements as some trajectories are drawn outside the physically delimited track bounds, where pedestrians would be unable to reach.

Although the location system used can, under ideal conditions, achieve an accuracy of up to 20 cm, experiments and test scenarios could only reach sub-meter accuracy (Xavier 2011). The inaccuracies of the positioning system explain the irregularities in the trajectories, and can be attributed to imprecisions in the calibration process, limitations of the information filter used by the location system framework, background noise, the agglomeration of large number of tags in confined areas and the signal attenuation caused by the presence of a large number of test persons.

### 5.1. Analysis of the single file experiment

Only a straight section with length $l_m$ = 2 m was considered to determine the density-velocity relation of pedestrian movement. Entrance ($t^{en}$) and exit ($t^{ex}$) times were recorded for each pedestrian crossing the entrance ($x^{en}$) and exit ($x^{ex}$) of this section. From these times, both the average velocity

$$v_i = \frac{l_m}{t_i^{ex}-t_i^{en}} \quad (1)$$

of each crossing $i$ as well as the number of persons inside the measurement section $N(t)$ at each instant $t$ can be obtained. Taking into consideration the large period between consecutive measurements for each tag, a linear interpolation between the positions and instants when each tag is first located inside ($x_i^{in}$, $t_i^{in}$) or outside ($x_i^{out}$, $t_i^{out}$) the measurement section and the positions and times associated with the previous locations ($x_i^{in-1}$, $t_i^{in-1}$, $x_i^{out-1}$, $t_i^{out-1}$) are necessary to better estimate the exact entrance and exit times:

$$t_i^{en} = t_i^{in-1} + \left[(t_i^{in} - t_i^{in-1}) \frac{x^{en}-x_i^{in-1}}{x_i^{in}-x_i^{in-1}}\right] \quad (2)$$

$$t_i^{ex} = t_i^{out-1} + \left[(t_i^{out} - t_i^{out-1}) \frac{x^{ex}-x_i^{out-1}}{x_i^{out}-x_i^{out-1}}\right] \quad (3)$$

From the runs with a single pedestrian on the track, the free velocity vfree = 1.33 ± 0.13 m/s, was obtained, which matches well with the value from literature (1.34m/s) (Kuligowski 2010).

Density at each instant $t$ can be obtained from the instantaneous number of pedestrians $N$ inside the measurement section.

$$\rho(t) = N(t)/l_m \quad (4)$$

As the measurement section is short, only small numbers of persons can be inside. Consequently the value of density calculated from the above definition jumps between discrete values. An enhanced definition of the density, calculated through the time headways between successive pedestrians avoids this problem, but its calculation is a challenging task as the data collection system is unable to provide the location of two tags at the same instant, and also because inaccuracies would increase as error from two different measurements would have to be taken into account.

The density assigned to each pedestrian crossing the measurement section is determined as the mean value of density during the crossing:

$$\rho_i = \frac{1}{t_i^{ex}-t_i^{en}} \int_{t_i^{en}}^{t_i^{ex}} \rho_n dt \quad (5)$$

The differences between the mean value of density over time determined by this method or by the time headway method is negligible (Chattaraj 2009). Fig. 6 shows the evolution of the crossings' speed and density in the measurement section over the whole duration of the run with 20 pedestrians.

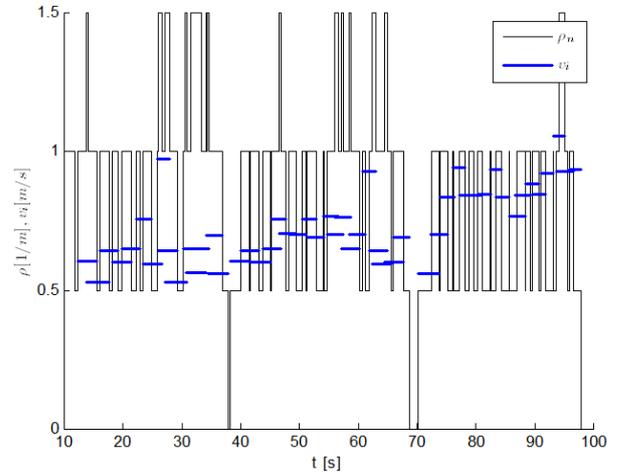

Figure 6: Evolution of speed $v_i$ and density $\rho_n$ over the duration of the experiment composed of 20 participants. Tick blue lines, whose length indicates the time interval a pedestrian is inside the measurement section, represent the mean velocity of the crossing.

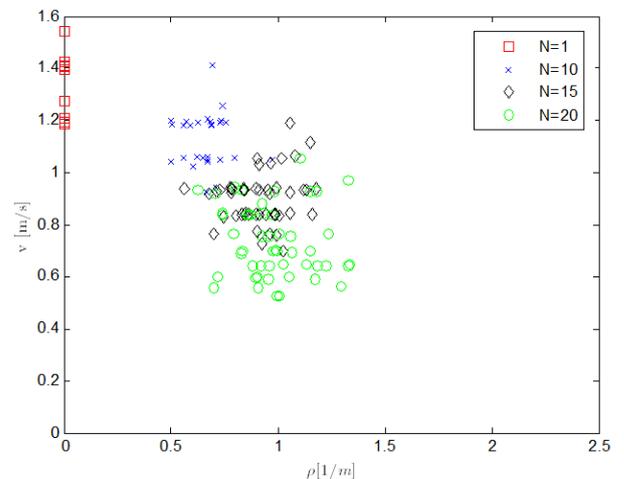

Figure 7: Relation between density and velocity in the single file scenario for the runs with 1, 10, 15 and 20 participants.

A graphical representation of the velocity ($v_i$) - density ($\rho_i$) pairs of each crossing is presented in Fig.7. This representation is known as the fundamental diagram of pedestrian movement.

In comparison with the diagrams obtained from similar experiments (Seyfried 2005, Chattaraj 2009), where data was collected manually from video recordings, similar values for density and velocity are found for runs with the same number of participants. However, our diagram is more disperse as a result of the limited precision of the UWB system. No data is presented for values of density over 1.2 1/m because when the experiment was performed no more than 20 voluntaries were present.

## 6. CONCLUSIONS

UWB based systems present a great potential for pedestrian data collection. Although more experiments are needed to validate this technique, it shows potential and suitability for human trajectory extraction.

Usage of this technology presents several advantages over traditional data collection techniques, expanding the breath of possible scenarios for experiments, such as situations of limited visibility for which data is inexistent. Compared with video recordings for tracking of people trajectories, UWB allows use in narrower spaces, lower ceilings and areas with line of sight restrictions. Other aspect is related with data collection. UWB based systems record the coordinates of position for each tag directly whereas video recordings must be later analysed and processed in order to extract positions/trajectories. Each tag carried by pedestrians is uniquely identifiable, allowing individual tracking thus enabling investigating both the dynamics of crowds and study the behaviour of specific individuals like child, elderly or people with mobility impairments.

On the down side, it has a lower sample rate than video: about 5 Hz in the case of UWB versus 25 Hz with video. Also, video techniques present synchronized results whilst UWB does not. Moreover, technical limitations such as lack of fine spatial precision make it not optimal for extracting microscopic properties of traffic, such as velocities and densities at a disaggregated level.

In conclusion, and when comparing the drawbacks with the advantages, UWB techniques for human trajectory extraction seems a viable approach. It is particularly suitable for scenarios where video is less applicable and pedestrian fine positioning is not an issue, such as for macroscopic analysis (e.g. egress time, path choice and behaviour scrutiny).

Future work will focus on the analysis of other experiments using different scenarios, following the same methodology. The results of this work will be used to validate and calibrate behaviour models in the "mSPEED" framework. This will constitute a unique tool for agent-based "Modelling and Simulation of Pedestrian Emergent Evacuation Dynamics" under development at LIACC.

## ACKNOWLEDGMENTS
Authors would like to thank the volunteers who participated in this study. The second and third authors also acknowledge the financial support from FTC (Fundação para a Ciência e a Tecnologia) in form of PhD scholarships (grants SFRH/BD/60727/2009 and SFRH/BD/72946/2010).